\documentclass[a4paper]{article}

\usepackage{mathpazo}
\usepackage[T1]{fontenc}
\usepackage[utf8]{inputenc}

\usepackage{amsmath}
\usepackage{amsfonts}
\usepackage{amssymb}
\usepackage{amsthm}

\usepackage{tikz}
\usetikzlibrary{positioning}

\title{How Perfect Offline Wallets Can Still Leak Bitcoin Private Keys}
\author{Stephan Verbücheln\footnote{Humboldt-Universität zu Berlin}}

\begin{document}

\maketitle

\begin{abstract}
ECDSA has become a popular choice as lightweight alternative to RSA and classic DL based signature algorithms in recent years. As standardized, the signature produced by ECDSA for a pair of a message and a key is not deterministic. This work shows how this non-deterministic choice can be exploited by an attacker to leak private information \textit{through} the signature without any side channels, an attack first discovered by Young and Yung for classic DL-based cryptosystems in 1997, and how this attack affects the application of ECDSA in the Bitcoin protocol.
\end{abstract}

\section{Introduction}
Bitcoin is decentralized payment scheme first described in a publication by pseudonymous author Satoshi Nakamoto in 2008.\cite{nbtc} Because its design is based on cryptographic protocols, the term \textit{cryptocurrency} has become common to refer to systems like Bitcoin and its variants. In a nutshell, users hold public-private key pairs, where the public keys function as an account numbers, and the private keys enable them to spend money from those accounts.

Bitcoin has become very popular and commercially adopted in recent years. The exchange rate has risen to a point that the value of all coins in circulation adds up to \$5 billion as of November 2014 (peaking even higher before). Because of the value of Bitcoin assets, it has become more and more important for users -- especially large-scale users like online shops -- to protect their private keys, as stealing a Bitcoin private key enables an adversary to steal their money.

Many ideas have come up on how to keep the user's private keys secret. The regular Bitcoin wallet software for PC operating systems encrypts the private keys using AES and only decrypts them when the user wants to create a transaction. This gives basic protection against PC malware for the time the user does not use his private key. But when the user wants to generate a transaction and enters his passphrase, the unencrypted keys are in the application memory, and thus are not protected against malware.

The next step is to use a dedicated PC for Bitcoin only, and keep it disconnected from the Internet. This prevents the PC from being infected with malware, and even in case it gets infected, the malware has a hard job to send the secret keys home to the attacker. As an even further step, some companies start producing special purpose hardware which is designed to manage keys and create transactions, but to never release the keys themselves.

The strength of the latter two settings comes from the fact that even if the device with the Bitcoin private keys on it is malicious, it cannot send any private information to attackers as long as the user makes sure that only transactions leave the device. This is supposedly achieved, if the device for example writes the transaction into a file, and the user copies the file using a portable storage medium.

But, as we will see later, even if the user actually ensures that only the transaction leaves the wallet device, there is still a problem: ECDSA. Bitcoin transactions contain ECDSA signatures. In the process of creating a ECDSA signature, like in the classic DSA, the creator has to choose a random number. Adam Young and Moti Yung showed in \cite{yy}, how this random number can be used by a malicious implementation of DSA to leak private information. This can be done in such a way that only the attacker, who made the malicious implementation, can extract the secret from the signature, and that the distribution of malicious signatures is polynomially indistinguishable from legitimate signatures by anybody else.

This paper will outline how this attack can be used for ECDSA as well, and what security issues arise from that for applications of ECDSA like Bitcoin.

\section{Definitions}
In our considerations, the \textit{user} is the person, who wants to use a publicly specified cryptosystem, e.g. ECDSA. The \textit{attacker} is the person, who creates a malicious implementation which is used by the user. The intuition is that the attacker's impementation can differ from how the cryptosystem is specified, but that the inputs and outputs have to comply with the specification. In addition we don't want the user or any third party to be able to distinguish the outputs generated by the malicious implementation from the outputs generated by a specification-compliant implementation.

\subsection{Kleptographic SETUP}
We will use the definition of kleptographic setups given by Young and Yung in \cite{yy}.

\paragraph{Definition.} Let $C$ be the black-box implementation of a cryptosystem with publicly known specification. A kleptographic \textit{(regular) SETUP} (Secretly Embedded Trapdoor with Embedded Protection) is a modified algorithm $C'$ such that:
\begin{enumerate}
\item The input of $C'$ agrees with he public specifications of the input of $C$.
\item $C'$ computes efficiently using the attacker's public encryption function $E$ (and possibly other functions as well), contained within $C'$.
\item The attackers private decryption function $D$ is not contained within $C'$ and is known only by the attacker.
\item The output of $C'$ agrees with the public specifications of the output of $C$. At the same time, it contains published bits (of the user's secret key) which are easily derivable by the attacker (the output can be generated during key generation or during system operation like message sending).
\item Furthermore, the output of $C$ and $C'$ are polynomially indistinguishable (as in \cite{gm84}) to everyone except the attacker.
\item After the discovery of the specifics of the setup algorithm and after discovering its presence in the implementation (e.g. reverse-engineering of hardware tamper-proof device), users (except the attacker) cannot determine past (or future) keys.
\end{enumerate}

\section{A Kleptogram for Elliptic Curves}
In this section, we will see how elliptic curves can be used to hide information in the choice of a random number like it is performed in ECDSA. This will be the central building block in our later attack.

Given an elliptic curve $E$. Let $G$ be a point on $E$ of order $n$. Let $d$ be the attacker's private key and $Q = dG$ the corresponding public key. Let $\mathcal{R}$ be a cryptographically strong pseudo-random number generator with hidden seed. Without loss of generality, we assime that $\mathcal{R}$ outputs a value less than $n$.
Let $\alpha, \beta, \omega$ be fixed integer constants, with $\omega$ being odd.

\subsection{Generation}
The malicious implementation of a elliptic-curve cryptosystem generates two subsequent choices $c_1, c_2$ the following way:
\paragraph{First round.}
\begin{enumerate}
\item Pick random $c_1 < n$.
\item Compute $M_1 = c_1 G$.
\item Store $c_1$ in non-volatile memory.
\item Output $M_1$.
\end{enumerate}
\paragraph{Second round.}
\begin{enumerate}
\item Pick random bit $t \in \{0,1\}$.
\item Compute $Z = (c_1 - \omega t) G \ + \ (-\alpha c_1 - \beta) Q$.
\item Compute $c_2 = \mathcal{R}(Z)$.
\item Compute $M_2 = c_2 G$.
\item Output $M_2$.
\end{enumerate}

\subsection{Recovery}
Now the attacker is able to extract the (secret) value of $c_2$ from $M_1, M_2$ and his private key $d$ as follows:
\begin{enumerate}
\item Compute $R = \alpha M_1 + \beta G$.
\item Compute $Z_1 = M_1 - d R$.
\item If $M_2 = \mathcal{R}(Z_1) G$ then output $c_2 = \mathcal{R}(Z_1)$.
\item Compute $Z_2 = Z_1 - \omega G$.
\item If $M_2 = \mathcal{R}(Z_2) G$ then output $c_2 = \mathcal{R}(Z_2)$.
\end{enumerate}

\subsection{Indistinguishability}
Values in the first round are chosen at random. An attacker wants the distribution of values from the second round to be indistinguishable from values generated at random like in the first round.

For this reason, the second round makes use of a seeded pseudo-random number generator. To make the distribution of the generator's output as hard to distinguish as possible, the attacker wants the number of potential values for $Z$ as large as possible, ideally $n$ different values.

The attacker can achieve this by tweaking the constants $\alpha$, $\beta$ and $\omega$ such that
\begin{align*}
G_1 &= (-d\beta - \omega)G \\
G_2 &= (-d\beta)G \\
G_3 &= (1 - d\alpha)G
\end{align*}
have preferably high orders, the ideal being $n$. In that case, $Z$ and thus $c_2 = \mathcal{R}(Z)$ can take on any value up to $n$. Assuming that the output of $\mathcal{R}$ is polynomially indistinguishable from random numbers of equal distribution, we can follow that only the attacker himself can distinguish the two (by blindly applying his recovery procedure).

\section{ECDSA and Bitcoin}
In this section, we will see how the above construction can be used to create a SETUP for ECDSA.

\subsection{Setup for ECDSA}
Let's recall ECDSA. For an elliptic curve $E$ with a generator point $G$ of order $n$, an ECDSA user's private key is a number $d < n$. The public key is the corresponding curve point $Q = dG$. Let $\mathcal{H}$ be the cryptographic hash function used.

A signature for a message $m$ is generated as follows:
\begin{enumerate}
\item Pick a random value $k < n$.
\item Compute $R = (r_1,r_2) = kG$.
\item Compute $r = r_1 \mod n$.
\item Compute $s = k^{-1}(\mathcal{H}(m) + dr) \mod n$.
\item Output $\sigma = (r,s)$
\end{enumerate}
The signature is verified by another party as follows:
\begin{enumerate}
\item Compute $R' = (r'_1, r'_2) = s^{-1}\mathcal{H}(m)G + s^{-1}rQ$.
\item Compute $r' = r'_1 \mod n$.
\item Accept the signature if and only if $r = r'$.
\end{enumerate}

Note that a number $k_i$ is chosen by the algorithm for each signature $\sigma_i$. A black-box implementation of ECDSA can substitute the values $k_i$, $k_{i+1}$ in two consecutive signatures $\sigma_i$, $\sigma_{i+1}$ by $c_1$, $c_2$ generated like in section 3. This enables the attacker to extract the value of $k_{i+1} = c_2$ from $\sigma_1 = (r_1, s_1)$, $\sigma_2 = (r_2, s_2)$ as described, because $r_1,r_2$ are the equivalent to $m_1,m_2$ respectively from section 3.

Now note that knowledge of the value $k$ for a single signature $\sigma$ already enables an attacker to compute the private key $d$, since from step 4 in the signature generation, it follows:
\[
d = (\mathcal{H}(m) - sk)r^{-1} \mod n
\]

\subsection{Limitations}
One limitation with this attack is that the attacker has to know in advance which curve and which generator point the user will choose to set up his implementation. Otherwise he cannot generate his own public-private key pair. This was a big limitation, when \cite{yy} published the attack for classic Diffie-Hellman settings, because it was common that each user generates his own new group $\mathbb{Z}_p$ and a generator $g \in \mathbb{Z}_p^*$ for each application.

With elliptic curve cryptography, this limitation became less relevant. Since it is more expensive to generate a new strong pair of a curve and a generator point from scratch, elliptic curves are generated and analyzed by scientists, and only a few elliptic curves are published (with corresponding parameters including the generator point) by standardization bodies such as the U.S. governments' NIST. Some ECDSA applications even specify a particular single curve to be used, e.g. Bitcoin uses only the secp256k1\footnote{Published in 2000 by Certicom Research.} curve for ECDSA.

\subsection{Bitcoin}
\begin{figure}
\begin{center}
\begin{tikzpicture}[box/.style={draw,text width=2.5cm,align=center}]
\node[box] (wal) {Offline Wallet};
\node[box, above=of wal] (pc) {PC/Server};
\node[above=of pc] (out) {Outside World};
\draw[<->] (out) -- node[auto] {general communication} (pc);
\draw[<->] (pc) -- node[auto] {requests, signed transactions} (wal);
\end{tikzpicture}
\end{center}
\caption{Bitcoin configuration with an offline wallet}
\end{figure}
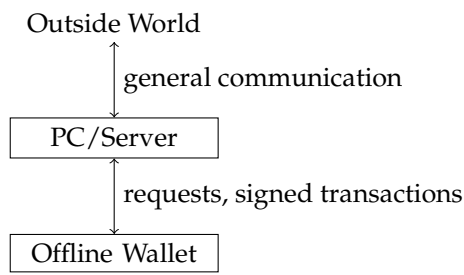
Bitcoin is a decentralized payment system, which is implemented as a peer-to-peer network.\cite{nbtc} In order to pay money to one another, users create little files named \textit{transactions}. These transactions are broadcasted into the network and collected in a large log file called \textit{blockchain}, which is maintained by all participating peers in a cooperative manner.

For our considerations we don't have to go further into how the blockchain is maintained in detail. We just have to keep in mind that the blockchain collects all valid transactions, and that the previous blockchain history is what gives the user a balance that he can spend.

To participate in the payment and receive money, a user creates a Bitcoin \textit{address}.
He creates an ECDSA public-private key pair.\footnote{Note that the secp256k1 curve is the only elliptic curve used in Bitcoin.} To generate the address, the public key is hashed twice using the SHA256 algorithm.

Now a standard transaction is a file that specifies a number of inputs and outputs. For inputs, the transaction refers to a previous transaction that has charged one or more of the user's addresses with coins. For outputs, the transaction specifies one or more Bitcoin addresses owned by the receiver. To make the transaction valid, it has to satisfy the following conditions:

\begin{itemize}
\item The specified inputs must have a higher or equal value than the outputs are receiving.
\item The transactions specified in the inputs must be part of the blockchain and not yet spent.
\item The user has to prove that he owns the addresses from the input transactions by signing the transaction with the corresponding private keys.
\end{itemize}
The last condition is the one that makes sure that only the owner of an address can spend the received coins, because only he knows the ECDSA private key.

Now note that since the signatures in a transaction are plain ECDSA signatures, the kleptographic SETUP from section 4 can just be applied in a straight forward manner. This means that a malicious programmer or hardware designer can implement a Bitcoin wallet in a way that leaks the secret key without any side channels using only two signatures. Note that since there might be more than one signature in a transaction, this can happen in a single transaction if two inputs associated with the same address are used. Nobody but the attacker is able to distinguish such a malicious transaction from a normal one.

\subsection{Deterministic Choice of $k$}
The kleptographic attack on ECDSA is very easy, because the value $k$ has to be chosen during the signature creation. As we have seen, $k$ has to be secret to prevent adversaries from extracting the secret key $d$. Only $R = kG$ is published with the signature.

Another slightly related security issue also arose from the fact that $k$ has to be chosen by the signature algorithm. If two values $k_1$, $k_2$ in two different signatures have a known linear relationship $k_2 = a k_1 + b$ with $a, b \in \mathbb{Z}$, the private key $d$ can be extracted from the two signatures without the knowledge of the values $k_1, k_2$, since it results in two linear equations with only $d$ and $k_1$ unknown.

Because of this known attack, some proposals have been made on how to choose the number $k$ deterministically, e.g. \cite{rfc} or the specification of EdDSA by \cite{eddsa}. These proposals generate $k$ deterministically using a cryptographic hash function with the message and the user's private key as inputs. The result is that the same user always creates the same signature for a given message. Note that it is crucial here to have the private-key part of the input. If $k$ would be derived from the message alone, it would be public information and therefore useless.

This measure gives only limited protection against the kleptographic attack since it can only be verified using the private key. The whole point of a dedicated Bitcoin wallet is that the user wants to make sure that the private keys never get anywhere outside of it, which means that even the user himself cannot verify the signature using a second computer.

\section{Potential Solutions}
One potential counter-measure could be the following:
\begin{enumerate}
\item The signature device generates the signature with deterministic $k$ as specified in \cite{rfc}.
\item In addition to that, the device delivers a zero-knowledge proof that $k$ was indeed generated as specified.
\end{enumerate}
This is possible since it is known that zero-knowledge proofs exist for any NP statement. But with this solution, a new problem arises. The whole point of using an offline Bitcoin wallet is that it does not leak any information into the public except the legitimately generated transactions. If we let the device output a zero-knowledge proof in addition to that, this proof may introduce new ways for the adversary to leak information. This means that the zero-knowledge protocol for this application has to be chosen carefully.

Another counter-measure would be to strictly not use any address more often than once. Although this way of using Bitcoin addresses is recommended already because of privacy considerations, there exist some use cases where this may not be feasible. For example, a public donation address for a charitable foundation is supposed to be used by multiple donors over and over again for a long period of time. In order to transfer the donations, a signature has to be created for each incoming payment.

A deterministic choice of $k$ alone (as described in the last section) would not help much either, because knowledge of the private key is necessary to verify that $k$ has indeed been chosen as specified. But there are still two advantages arising from using a deterministic method: First, it limits the choice of $k$ if the user signs the exact same document twice, as $k$ has to be identical given the same message and private key. Secondly, it would enable the user to detect malicious signatures later, for example after a key has expired and it is safe to transfer the private key to another computer.

\subsection{Interactively Generate $k$}
The ability of the signer to leak information through his choice of $k$ come from the fact that he is allowed to choose $k$ freely. If an outside agent could force the signer to choose $k$ from the equal distribution, that wouldn't be possible. 

A common way to solve this cryptographic problem is an interactive protocol, where two parties choose a common random string. Such a protocol is arranged in a way that neither of the two parties can influence the resulting random string to his wishes, because both parties have to make their choice for their part before the other one's choice is revealed. This can be enforced using a commitment scheme.

In a blogpost\footnote{Blogpost from June 20, 2013, “No subliminal Channel”, http://firmcoin.com/?p=52} on firmcoin.com, the Certimix company describes such an interactive protocol that deals with the problem. In addition to the signing device (i.e. the offline wallet), we have a supervisor device that checks whether the signing is done properly. The protocol works as follows:

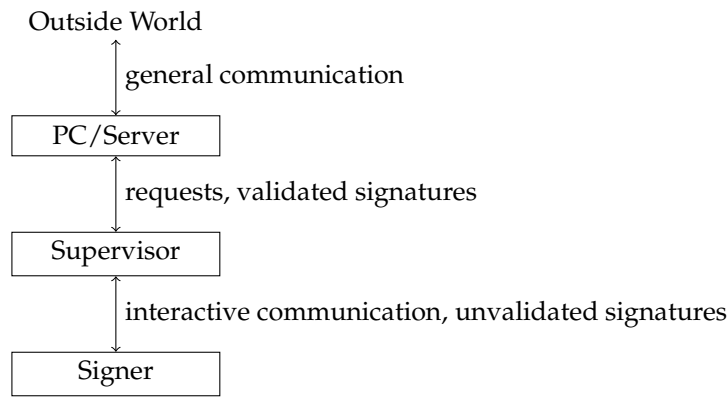
\begin{figure}
\begin{center}
\begin{tikzpicture}[box/.style={draw,text width=2.5cm,align=center}]
\node[box] (sup) {Supervisor};
\node[box, above=of sup] (pc) {PC/Server};
\node[above=of pc] (out) {Outside World};
\node[box, below=of sup] (sig) {Signer};
\draw[<->] (out) -- node[auto] {general communication} (pc);
\draw[<->] (pc) -- node[auto] {requests, validated signatures} (sup);
\draw[<->] (sup) -- node[auto] {interactive communication, unvalidated signatures} (sig);
\end{tikzpicture}
\end{center}
\caption{Configuration with a supervisor}
\end{figure}

For an elliptic curve $E$ with a generator point $G$ of order $n$ be $d < n$ the private key and the corresponding curve point $Q = dG$ the public key. Let $\mathcal{H}$ be the cryptographic hash function used.

A signature for a message $m$ is generated as follows:
\begin{enumerate}
\item Supervisor and signer generate $k$ together:
\begin{enumerate}
\item The signer picks a random value $t < n$.
\item The signer computes $T = tG$.
\item The signer computes $h_T = \mathcal{H}(T)$.\footnote{This is a commitment to $T$.}
\item The signer sends $h_T$ to the supervisor.
\item The supervisor picks a random value $u < n$.
\item The supervisor sends $u$ to the signer.
\item The signer sends $T$ to the user.
\item The supervisor verifies that $h_T = \mathcal{H}(T)$.
\item The signer computes $k = t \cdot u \mod n$.
\end{enumerate}
\item The signer computes $R = (r_1,r_2) = kG$.
\item The signer computes $r = r_1 \mod n$.
\item The signer computes $s = k^{-1}(\mathcal{H}(m) + dr) \mod n$.
\item The signer sends the resulting signature $\sigma = (r,s)$ to the supervisor.
\item The supervisor releases the signature after verifying that $k$ was generated correctly:
\begin{enumerate}
\item The supervisor computes $R' = (r_1',r_2') = uT$.
\item The supervisor verifies that $r = r'_1 \mod n$ holds.
\end{enumerate}
\end{enumerate}

Note that only the supervisor has to communicate with the outside world. But there are still some things that we have to be aware of: Even though neither the signer nor the supervisor can manipulate the choice of $k$/$r$ as published with the signature, the signer might still leak information to the supervisor through his choices of $t_i$. There are still advantages coming from the protocol: As the supervisor does not have to store any critical data in persistent memory, the security properties are different. In addition, as long as either one of signer and supervisor works correctly, the signature cannot leak information through the choice of $k$.

This protocol has the disadvantage that it requires a lot of interaction. But we can make this protocol more practical by performing a prearrangement step where a list of random numbers is generated in advance.

Let $E$ again be an elliptic curve with a generator point $G$ of order $n$, be $d < n$ the private key and the corresponding curve point $Q = dG$ the public key. Let $\mathcal{H}$ be the cryptographic hash function used.

\paragraph{Prearrangement}
\begin{enumerate}
\item The signer chooses a list of random numbers $t_1, \dots, t_\ell$ with $t_i < n$ for each $i \le \ell$.
\item For $i \le \ell$, the signer computes $T_i = t_i G$ and $h_{T_i} = \mathcal{H}(T_i)$.
\item The signer stores the lust of choices $t_1, \dots, t_\ell$ for later use.
\item The signer sends the list of hashes $h_{T_1}, \dots, h_{T_\ell}$ to the supervisor.
\item The supervisor stores the list of hashes.
\end{enumerate}

\paragraph{Signing}
\begin{enumerate}
\item The supervisor generates a random number $u < n$.
\item The supervisor sends $(m, u)$ to the signer.
\item The signer computes $k = u \dot t_1 \mod n$ and $T_1 = t_1 G$, where $t_1$ is the first element from the list of choices.
\item The signer generates the ECDSA signature $\sigma = (r,s)$ using $k$ as the nonce.
\item The signer sends $(\sigma, T_1)$ to the supervisor and removes $t_1$ from the list of choices.
\item The supervisor computes $R' = (r'_1, r'_2) = u T_1$.
\item The supervisor verifies that $r = r'_1 \mod n$ and $h_{T_1} = \mathcal{T_1}$.
\item If successfully verified, the supervisor publishes the signature and removes $h_{T_1}$ from the list of hashes.
\end{enumerate}
If any verification step fails, the supervisor should cancel the protocol and alert the user.

\section{Conclusion}
Without a satisfying solution, there is only one conclusion to draw from this problem: Users cannot trust any implementation of ECDSA or Bitcoin, which they cannot fully verify.

Note that this does not only affect strict black-box implementations such as closed-source programs. A user with high security requirements (like an online shop that wants to accept Bitcoin payments on a large scale) would have to use an implementation, which can be verified by his own staff (or at least a reliable partner company). An hard-to-read implementation like OpenSSL may be insufficient, because the particular implementation of ECDSA may be hard to verify. Such implementations feature many variations on the pure algorithm to improve performance (e.g. using CPU-dependent assembly language) or to harden the implementation against timing attacks). In addition, it is hard to verify in a large program like OpenSSL, which code is actually executed when you perform a certain operation.

A similar problem arises with embedded cryptographic chips like smartcards. Such devices are designed to never release the private keys and to make it hard for an outside analysis to read out secret data. The fact that the leak in the kleptographic attack is so well hidden makes it hard for a chip manufacturer to prove to the customers that the device does not leak any secret information in ECDSA signatures.

The paranoid among users would even have to compile the program themselves to be sure that the code their are reading really matches the code they are running. To verify that the executed code actually matches the source code is even harder for small embedded devices (like dedicated Bitcoin wallets or crpyto smartcards) than in the setting of an offline PC.

\end{document}